\documentclass[twocolumn,prl,secnumarabic,amssymb,amsmath,nobibnotes,aps,showpacs]{revtex4-2}
\usepackage{graphics}
\usepackage{graphicx}
\usepackage{amssymb}
\usepackage{braket}
\usepackage{float}
\usepackage[usenames,dvipsnames]{color}
\usepackage{mathrsfs}
\usepackage{unicode}
\usepackage{ifsym}

\begin{document}

\title{Resonance Raman optical cycling for high-fidelity fluorescence detection of molecules}

\author{J.~C. Shaw$^{\dagger}$, J.~C. Schnaubelt \& D.~J. McCarron$^{*}$}
\affiliation{Department of Physics, University of Connecticut, 196A Auditorium Road, Unit 3046, Storrs, Connecticut 06269, USA.}

\begin{abstract}
We propose and demonstrate a novel technique that combines Raman scattering and optical cycling in molecules with diagonal Franck-Condon factors. This resonance Raman optical cycling manipulates molecules to behave like efficient fluorophores with discrete absorption and emission profiles that are readily separated for sensitive fluorescence detection in high background light environments. Using a molecular beam of our test species, SrF, we realize up to an average of $\approx20$ spontaneously emitted photons per molecule, limited by the interaction time, while using a bandpass filter to suppress detected scattered laser light by $\sim10^{6}$. This general technique represents a powerful tool for high-fidelity fluorescence detection of molecules in any setting and is particularly well-suited to molecular laser cooling and trapping experiments.
\end{abstract}

$\pacs{}$

\maketitle
Ultracold molecules provide a versatile platform for tests of fundamental physics and strongly interacting many-body systems \cite{Carr2009}. Recent progress directly cooling and trapping molecules at ultracold temperatures \cite{Prehn2016,Norrgard2016,McCarron2018} has realized a new path towards the full quantum control of a diverse range of species. This diversity is favorable for applications including the advancement of ultracold chemistry \cite{Hu2019} and improved precision measurements \cite{Truppe2013,Hutzler2020}. High-fidelity detection of ultracold molecules is central to almost all of their proposed applications and is a growing area of research \cite{Zeppenfeld2017,Cheuk2018,Guan2020}. Popular sensitive detection methods for molecules, such as resonance-enhanced multi photon ionization (REMPI) and fluorescence detection using photomultiplier tubes, lack the spatial-information necessary to extract often-essential properties such as position, density, lattice filling fraction or optical tweezer occupancy. To-date, the densities of directly cooled samples of molecules are several orders of magnitude too low for absorption imaging to be viable \cite{McCarron2018b} and this technique has been limited to dense samples of molecules assembled from pairs of laser-cooled atoms \cite{Wang2010}.

Imaging fluorescence onto a charge-coupled device (CCD) camera is a powerful method for detecting molecules with spatial-information and is a natural choice for laser-coolable species with optical cycling transitions \cite{Shuman2009}. However, today's multilevel optical cycling schemes in molecules demand high laser intensities for laser cooling and trapping \cite{Barry2014}, while absorbing and emitting photons at the same wavelength, which results in scattered laser light and its associated shot noise being the dominant imaging noise source. Until now, scattered light has been managed using blackened vacuum chambers that avoid scattering surfaces near the molecules but this restricts optical access \cite{Norrgard2016b}. Many future applications for ultracold molecules will likely require improved optical access and the introduction of scattering surfaces nearby, such as chip-based microtrap arrays \cite{Meek2009b,Santambrogio2015,Hou2017} or superconducting circuits \cite{Andre2006}, and an alternative solution is required to mitigate the effects of scattered light to achieve high-fidelity fluorescence detection.

Here, we propose and demonstrate a novel fluorescence imaging technique, inspired by the use of fluorophores in the life-sciences \cite{Chalfie1994,Lakowicz1999}, called resonance Raman optical cycling (RROC). Our approach combines Raman scattering with optical cycling to enable laser-coolable molecules to spontaneously emit multiple photons which are frequency-shifted away from the driving laser fields. This enables high-fidelity separation of scattered laser light and spontaneous emission for detection, much like the Stokes shift leveraged from fluorophores which is derived from vibrational relaxation of these typically large and complex molecules. Using this technique we are able to suppress the detected scattered light background by $\sim10^{6}$, realizing a platform for sensitive fluorescence detection of single molecules in high background light environments without its associated shot noise. Our analysis compares the expected photon scattering rate for RROC against the typical optical cycling approach \cite{Shuman2009} and provides a simple connecting relationship. We demonstrate the versatility of this technique by using RROC to measure the forward velocity distribution of our SrF molecular beam. Finally, we discuss other relevant molecules for which this method is applicable.

The general technique of RROC requires two independent lasers applied to molecules with near-diagonal Franck-Condon factors (FCFs) [Fig. 1(a)]. First, a Stokes laser excites population from the electronic ground state with vibrational quantum number $v=0$ to an electronic excited state with $v$'$=1$. Molecules then spontaneously decay with near unit probability into $v=1$ within the electronic ground state and spontaneously emit photons shifted \emph{down} in energy by $\approx\omega_{e}$ compared to the Stokes laser (where $\omega_{e}$ is the ground state vibrational constant). Subsequently, an anti-Stokes laser excites population from $v=1$ in the electronic ground state into $v$'$=0$ within the electronic excited state. Molecules then return with high probability back into the original $v=0$ electronic ground state via spontaneous emission that is shifted \emph{up} in energy by $\approx\omega_{e}$ relative to the anti-Stokes laser.

\begin{figure}[t!]
  \centering
      \includegraphics[width=0.48
     \textwidth]{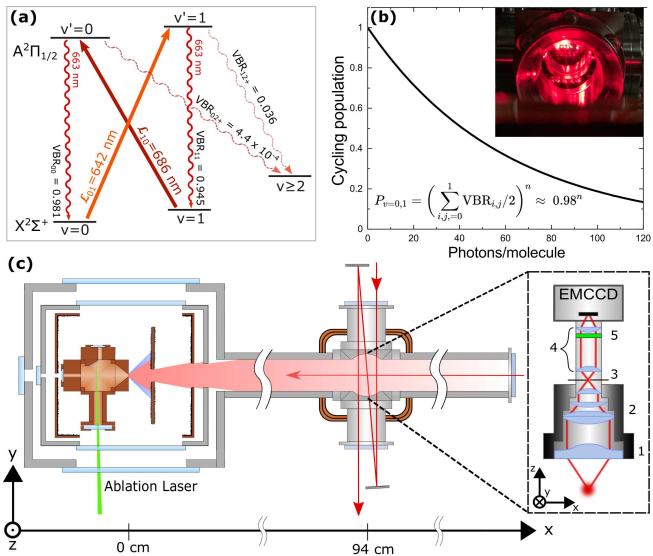}
  \caption{(a) Relevant energy levels for RROC in SrF addressed by the $\mathscr{L}_{01}$ and $\mathscr{L}_{10}$ lasers.  Diagonal vibrational branching ratios (VBRs) ensure that $>95~\%$ of LIF from both excited states is at 663~nm. (b) RROC population versus number of spontaneously emitted photons per molecule. SrF molecules have a $0.98$ average probability of remaining in the RROC system following spontaneous emission. Inset, photograph of the illuminated imaging region with no measures to reduced scattered laser light. (c) Schematic of the experimental setup showing the cryogenic source, imaging region and camera optical setup with 1. condenser lens, 2. commercial  50~mm $f\textbackslash0.95$ lens, 3. iris, 4. $1:1$ telescope and 5. bandpass filter.}
  \label{fig1}
\end{figure}

Alternating excitation of molecules by the Stokes and anti-Stokes lasers can continue in this system until a dark vibrational ground state with $v>1$ is populated, provided that rotationally closed transitions are used and dark sublevels are continuously remixed \cite{Stuhl08}. For laser-coolable molecules, this can correspond to $\sim50$ cycles around the four-level system and $\sim100$ spontaneously emitted photons per molecule that are readily isolated from the Stokes and anti-Stokes excitation photons using a bandpass filter [Fig. 1(b)]. Until RROC, generating Stokes-shifted fluorescence from molecules with diagonal FCFs resulted in only one emitted photon per molecule \cite{Galica2018}.

Our experimental setup uses a cryogenic buffer gas beam source, described elsewhere \cite{Shaw2020}, to produce cold, slow pulses of our test-species SrF containing $\sim10^{11}$ molecules per steradian per pulse in the $X^{2}\Sigma\ket{v=0}$ state. Molecules are detected by imaging laser-induced fluorescence (LIF) onto an electron multiplying CCD (EMCCD) $\approx1~$m downstream from the source. Up to three transverse passes of a laser beam below the camera excite molecules with light from a Stokes laser ($\mathscr{L}_{01}$) at 642~nm and an anti-Stokes laser ($\mathscr{L}_{10}$) at 686~nm [Fig. 1(c)]. Here, $\mathscr{L}_{vv^{\rm{'}}}$ denotes a laser tuned to the $X^{2}\Sigma^{+}\ket{v,N^{P}=1^{-}}$ $\rightarrow$ $A^{2}\Pi_{1/2}\ket{v^{\rm{'}},J^{\rm{'}\emph{P}^{\rm{'}}}=1/2^{+}}$ transition, where N is the angular momentum excluding spin, $\vec{J}=\vec{N}\pm\vec{S}$ (where $S=1/2$ is the electron spin), $P$ is the parity, and prime indicates the excited state. Radio-frequency (R.F.) sidebands are added to each laser to address the resolved ground state spin-rotation and hyperfine structure, and dark ground state sublevels are remixed using a 2~G magnetic field applied at $45^{\circ}$ to the linear polarizations of the $\mathscr{L}_{01}$ and $\mathscr{L}_{10}$ lasers. Following excitation, diagonal FCFs in SrF ensure that almost all LIF is at 663~nm and no measures are taken to reduce scattered laser light [Fig. 1(b) inset].

The imaging optics include a 15~nm FWHM bandpass filter centered at 662~nm to transmit LIF to the camera while blocking scattered laser light. A 1:1 telescope ensures that rays make an angle-of-incidence with the filter $\lesssim 15^{\circ}$ for efficient transmission of LIF and effective blocking of  Stokes laser light $\mathscr{L}_{01}$; an iris, placed before the telescope limits spherical abberations. As a final precaution, the inside walls of the imaging assembly are lined with black felt to reduce residual internal reflections. We measure the imaging system to suppress scattered laser light at 642 and 686~nm, relative to LIF at 663~nm, by factors of $10^{5}$ and $10^{7}$ for $\mathscr{L}_{01}$ and $\mathscr{L}_{10}$, respectively. Given the variety of imaging conditions presented in the literature, we use normalized units to compare the scattered light background signal in our system against other experiments and typically detect 40~photons/sec/mm$^{2}$/mW. To the best of our knowledge, this is a factor of 2 below the lowest value previously reported in a molecular laser cooling experiment, which used a blackened vacuum chamber \cite{Williams2017}.

\begin{figure}[h t b]
\hspace*{-0.5cm}
  \centering
      \includegraphics[width=0.48
     \textwidth]{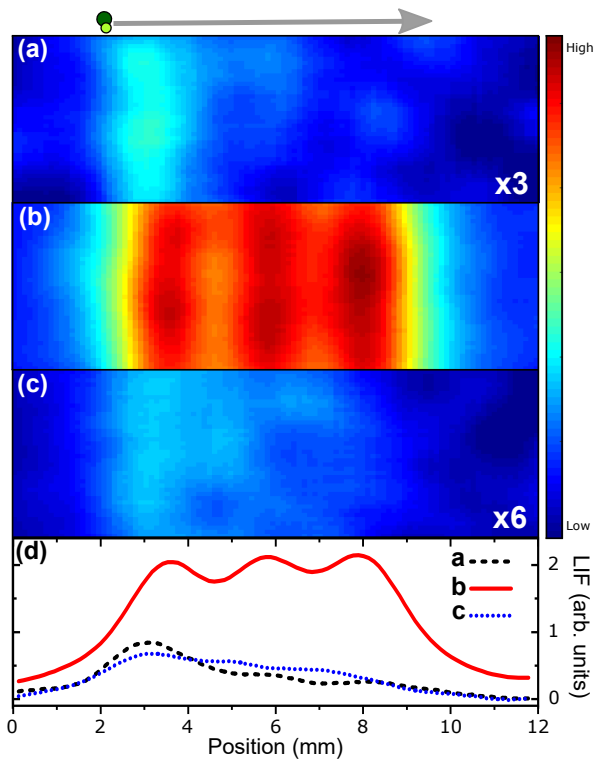}
  \caption{Single-shot background subtracted fluorescence images recorded for (a) $\mathscr{L}_{01}$ alone, (b) $\mathscr{L}_{01}$ and $\mathscr{L}_{10}$, and (c) $\mathscr{L}_{00}$. In each case, on average, molecules spontaneously emit 1, 13 and $\approx60$ photons, respectively, as they travel from left to right during a 30~ms exposure. (d) Integrated horizonal LIF profiles highlight the $13\times$ LIF increase $\epsilon$ between (a) and (b) and show evidence of molecules being pumped into dark states for case (a) and (c). For clarity, the LIF images and profiles for (a) and (c) have been multiplied by factors $3$ and $6$, respectively.}
  \label{fig2}
\end{figure}

Molecules are initially probed and imaged using 34~mW of $\mathscr{L}_{01}$ alone, which rapidly pumps molecules into the dark $X^{2}\Sigma\ket{v=1,N=1}$ state and on average produces $1$ photon per molecule. Here, as expected, detected LIF decreases as the molecules propagate downstream during their $\approx50~\mu$s interaction time with the laser light [Fig. 2(a)]. The addition of 30~mW of $\mathscr{L}_{10}$ enables molecules to optically cycle via RROC, leading to an increase in detected LIF by a factor of $\epsilon\approx13$ compared to the $\mathscr{L}_{01}$ alone case [Fig. 2(b)]. This increase indicates that on average each molecule has scattered 13 photons, completing $\sim6$ loops around the RROC four level system, and there is no decrease in detected LIF as molecules move downstream. Here, the mean RROC photon scattering rate is approximated using $R_{\rm{sc}}=13/50~\mu$s$~\approx3\times10^{5}~$s$^{-1}$, roughly an order of magnitude less than typical photon scattering rates for molecules \cite{Norrgard2016}. In this case, $95~\%$ of detected photons are due to LIF. By applying $\mathscr{L}_{10}$ alone we probe for molecules originally in the $X^{2}\Sigma\ket{v=1,N=1}$ state and detect no LIF in this configuration. This shows that the increased LIF detected when adding $\mathscr{L}_{10}$ to $\mathscr{L}_{01}$ is due to RROC and not to exciting previously dark molecules originally in $X^{2}\Sigma\ket{v=1,N=1}$. The largest RROC LIF increase measured during this work was $\epsilon\approx20$.

To compare the typical optical cycling approach \cite{Shuman2009} against RROC, we excite molecules with $7$~mW of light from a $\mathscr{L}_{00}$ laser at 663~nm. Now, scattered laser light and LIF are at the same wavelength and both are transmitted by our imaging system. Here, detected LIF decreases as molecules travel downstream and are optically pumped into $X^{2}\Sigma\ket{v=1,N=1}$, after each molecule spontaneously emits $\approx60$ photons [Fig. 2(c)]. In this case, an increased scattered light background results in only $1~\%$ of detected photons being due to LIF, despite a 60/13$\sim$4.6 fold increase in the average number of photons emitted per molecule vs RROC. A $150\times$ decrease in EMCCD gain was also necessary to avoid saturation. We highlight that all three LIF images in Fig. 2 are single $30$~ms exposures.

\begin{figure}[h t b]
\hspace*{-0.3cm}
  \centering
      \includegraphics[width=0.48
     \textwidth]{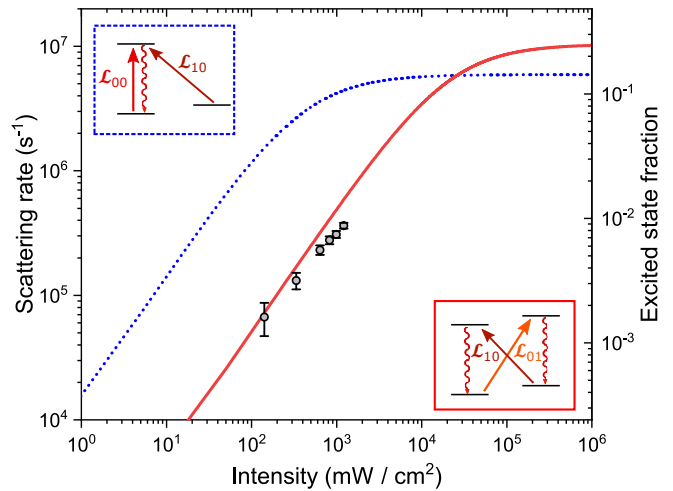}
  \caption{Calculated scattering rate versus laser intensity for the typical optical cycling approach (\textcolor{blue}{\textbf{$\cdots$}}) and RROC (\textcolor{red}{\textbf{---}}) alongside measured RROC scattering rates (\textcolor{Gray}{\textbf{$\bullet$}}). Below saturation, for fixed laser intensities the calculated RROC scattering rate in SrF is $\approx27\times$ smaller than that for the standard approach, in reasonable agreement with our measurements.}
  \label{fig3}
\end{figure}

We use multilevel rate equations \cite{Foot2007,Tarbutt2013} to model RROC, using $\mathscr{L}_{01}$ and $\mathscr{L}_{10}$ lasers, and compare this against the typical optical cycling approach, using $\mathscr{L}_{00}$ and $\mathscr{L}_{10}$ lasers to form a $\Lambda$-system. We assume equal intensities for the two lasers used in both systems. The maximum photon scattering rate in a multilevel system is $R^{\rm{max}}_{\rm{sc}}=\Gamma  N_{e}/(N_{g}+N_{e})$, where $\Gamma$ is the natural linewidth and $N_{g}$ and $N_{e}$ are the number of ground and excited state sublevels, respectively \cite{Shuman2009}. For RROC, the maximum photon scattering rate is roughly twice that attainable in the $\Lambda$-system due to the increased number of excited state sublevels available. In practice, this increase is challenging to realize, due to high laser intensity demands, since the photon excitation rate $R_{\rm{ex}}\propto q_{v^{'}v}I$ \cite{Foot2007}, where $q_{v^{'}v}$ is the transition FCF and $I$ is the laser intensity, and effective RROC excites weak transitions with FCFs $\sim0.01$. For fixed laser intensities below saturation, the decrease in scattering rate when moving from the $\Lambda$-system to RROC can be approximated by $(q_{00}+q_{01})/(q_{01}+q_{10})$, which is a factor of $\approx27$ for SrF. We use the measured LIF increase, $\epsilon$, over a fixed $50~\mu$s interaction time to measure the RROC scattering rate as a function of laser intensity, while keeping the $\mathscr{L}_{01}$ and $\mathscr{L}_{10}$ intensities equal, and find reasonable agreement with the multilevel rate equation model [Fig. 3]. Our model also accurately predicts scattering rates reported by others using a $\Lambda$-system with SrF to within a factor $\sim2$ \cite{Norrgard2016,Barry2014}.

To demonstrate the versatility of RROC, we use a variant of this technique to measure the forward velocity distribution in our molecular beam. Here, we remove the R.F. sidebands from the $\mathscr{L}_{01}$ laser and choose to excite two non-rotationally closed transitions, $X^{2}\Sigma^{+}\ket{v=0,1,N^{P}=0^{+}}$ $\rightarrow$ $A^{2}\Pi_{1/2}\ket{v^{'}=1,0,J^{'P^{'}}=1/2^{-}}$, with $\mathscr{L}_{01}$ and $\mathscr{L}_{10}$, since the ground state hyperfine structure consists of only two levels spanning $\sim100~$MHz, which simplifies interpretation of the Doppler-shifted LIF profile.

\begin{figure}[b!]
  \centering
      \includegraphics[width=0.48
     \textwidth]{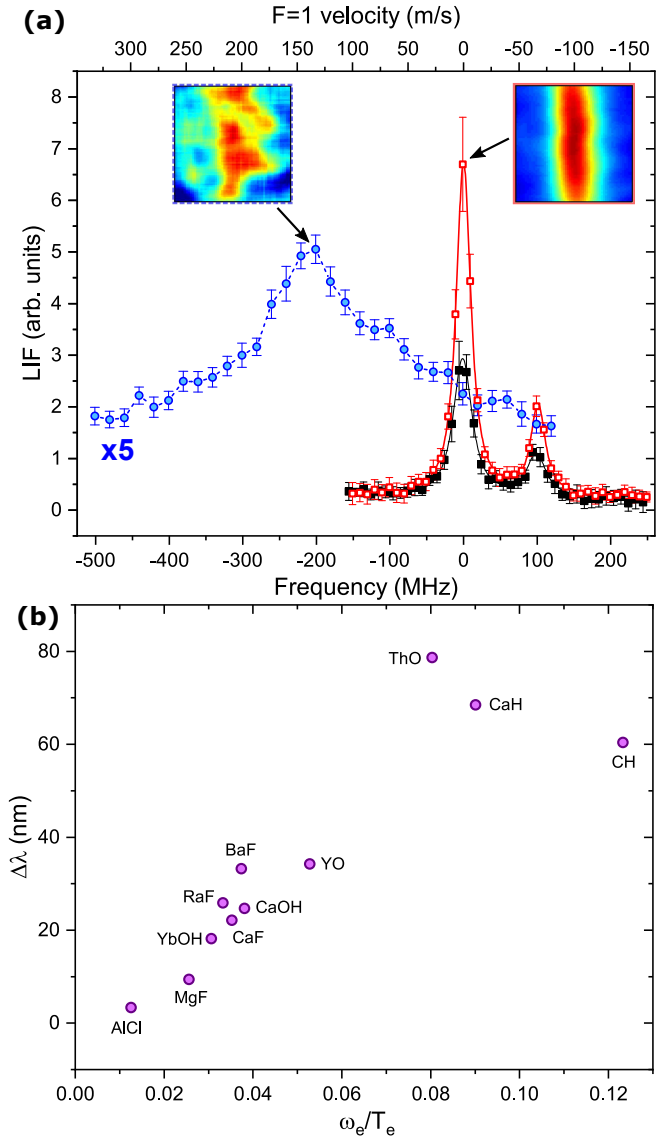}
  \caption{(a) Transverse (\textcolor{black}{$\blacksquare$} \textcolor{red}{$\square$}) and counterpropagating (\textcolor{blue}{$\bullet$}) LIF spectra and images, as described in the text, reveal the molecular beam forward velocity profile. LIF images for the transverse and counterpropagating cases are the average of $6$ and $24$ experimental cycles, respectively. Lorentzian fits are applied to the transverse data (\textcolor{black}{\textbf{---}} \textcolor{red}{\textbf{---}}) and the longitudinal line (\textcolor{blue}{\textbf{- - -}}) is to guide the eye. (b) Wavelength separation ($\Delta \lambda$) between emitted LIF and $\mathscr{L}_{01}$ versus $\omega_{e}/T_{e}$ for a selection of molecules studied today. RROC should be straightforward to implement for molecules with $\omega_{e}/T_{e}\gtrsim0.03$ since here $\Delta \lambda \gtrsim20$~nm.}
  \label{fig4}
\end{figure}

Initially we perform spectroscopy below the camera with negligible Doppler shift using a single transverse laser beam which, at first, contains only $\mathscr{L}_{01}$ [Fig. 4a \textcolor{black}{\textbf{$\blacksquare$}}]. This is followed by the same frequency scan of $\mathscr{L}_{01}$ but with $\mathscr{L}_{10}$ present and fixed on resonance to enable RROC [Fig. 4a \textcolor{red}{\textbf{$\square$}}]. The LIF increase measured between these two cases, $\epsilon$, reveals the average number of RROC photons scattered below the camera when addressing a specific hyperfine ground state; we measure 2.6 and 2.0 photons for the $F=1$ and $F=0$ ground states, respectively. The ratio of these values is dictated primarily by rotational branching into dark states and is close to the expected value of $\sim1.4$.

The single-frequency $\mathscr{L}_{01}$ laser is then applied counterpropagating to the molecular beam, while keeping the $\mathscr{L}_{10}$ laser transverse and below the camera, but now with three passes. Here, $\mathscr{L}_{01}$ rapidly shelves $2/3$ of the molecules Doppler-shifted onto resonance into $X^{2}\Sigma\ket{v=1,N=0}$ upstream of the detection region, before $\mathscr{L}_{10}$ acts to readout this shelved population for detection. The remaining $1/3$ of resonant molecules are lost into the dark $X^{2}\Sigma\ket{v=1,N=2}$ state due to rotational branching. Detected LIF as a function of the $\mathscr{L}_{01}$ frequency gives the forward velocity distribution in our molecular beam [Fig. 4a \textcolor{blue}{$\bullet$}]. Using the $F=1$ level as a reference, we determine a mean forward velocity of $\approx140$~m/s and a FWHM of $\approx70$~m/s, in agreement with previous measurements \cite{Shaw2020}. This approach is similar to that in Refs. \cite{Barry2012,Hemmerling2016}, which uses a ladder type two-photon transition to produce one UV photon per molecule. However, our RROC method can produce multiple photons per molecule and requires only one excited electronic state, which may be advantageous for polyatomic or short-lived radioactive molecules with little spectroscopic data available \cite{Augenbraun2020,GarciaRuiz2020}.

RROC is applicable to a large subset of molecules with diagonal FCFs. Implementing this technique can be straightforward when $\mathscr{L}_{01}$ and $\mathscr{L}_{10}$ for a given molecule differ in wavelength from the emitted LIF by $\Delta \lambda \gtrsim20$~nm and are readily separated using an off-the-shelf bandpass filter. This occurs when $\omega_{e}/T_{e}\gtrsim0.03$, where $T_{e}$ is the minimum electronic energy, and includes species actively studied today such as BaF \cite{Chen2017,Albrecht2020}, CaH \cite{Vazquez-Carson2021}, CaOH \cite{Baum2021}, CaOCH$_{3}$ \cite{Mitra2020}, CaF \cite{Jurgilas2021,Anderegg2021}, CH \cite{Schnaubelt2021}, OH \cite{Wcislo2021}, RaF \cite{GarciaRuiz2020}, SrF \cite{Langin2021}, ThO \cite{Wu2020}, YO \cite{Wu2021}, YbF \cite{Alauze2021} and YbOH \cite{Augenbraun2020} [Fig. 4b]. We note that for polyatomic species, the relevant vibrational constant $\omega_{e}$ refers to the symmetric stretching mode \cite{Baum2021}. RROC may also be possible for molecules with $\omega_{e}/T_{e}<0.03$, such as AlCl \cite{Daniel2021,Shaw2021}, AlF \cite{Hofsass2021} and MgF \cite{Meng2021} where $\Delta\lambda\approx3$, $4$ and $9$~nm, respectively, provided that custom bandpass filters are available. An alternative approach when $\omega_{e}/T_{e}<0.03$ could be the excitation of transitions with $\Delta v=2$ using $\mathscr{L}_{02}$ and $\mathscr{L}_{20}$ lasers. However, these transitions have even smaller FCFs and would demand a corresponding increase in the RROC laser intensities to maintain the same excitation rates.

Through realizing RROC in SrF, we have detected up to an average of $\approx20$ scattered photons per molecule in our molecular beam, limited by the $50~\mu$s interaction time, while suppressing background laser light signals by $\sim10^{6}$. Integrating RROC into existing laser cooling experiments for detecting trapped molecules would offer increased LIF signals compared to this work through significantly longer interaction times, and would simply require replacing the $\mathscr{L}_{00}$ laser for a $\mathscr{L}_{01}$ laser. The inclusion of a $\mathscr{L}_{21}$ repump laser would allow RROC to produce more photons per molecule before the dark $X^{2}\Sigma\ket{v=3,N=1}$ state is populated ($\sim2000$ photons for SrF) but would also decrease the photon scattering rate by introducing more ground state sublevels into the system. Including a $\mathscr{L}_{32}$ repump laser would allow even more photons to be scattered ($>10^{4}$ photons for SrF), with sufficient interaction time, and give no further decrease in the photon scattering rate. RROC realizes a platform for sensitive fluorescence detection of single molecules in high background light environments, similar to $\Lambda$-enhanced gray molasses cooling ($\Lambda$-cooling) used to image molecules in a blackened vacuum chamber where scattered laser light and LIF are at the same wavelength \cite{Cheuk2018,Anderegg2019}. We speculate that $\Lambda$-cooling could also be possible using RROC to cool and image trapped molecules with negligible detected scattered laser light, albeit at a reduced photon scattering rate.

An important consideration for RROC is the high laser intensities required to drive the weak optical transitions employed. In SrF we measure a maximum photon scattering rate that is $\sim10\times$ smaller than typical values using laser intensities that are $\sim10-20\times$ greater than the intensities commonly used in molecular MOTs beams \cite{Barry2014,Norrgard2016}. For trapped samples, higher RROC laser intensities (and photon scattering rates) could be realized using small imaging laser beams or large mode enhancement cavities \cite{Heinz2021} that specifically address the $\lesssim5~$mm$^{3}$ volume typically occupied by the molecules.

In summary, we have proposed and demonstrated a technique that combines Raman scattering and optical cycling for molecules with diagonal FCFs. This resonance Raman optical cycling (RROC) can manipulate molecules to behave like efficient fluorophores with discrete absorption and emission profiles that are readily separated for sensitive fluorescence detection in high background light environments. The production of Stokes-shifted fluorescence from fluorophores has been transformational within the life-sciences over the last century, leading to advances such as super-resolution spectroscopy \cite{Hell03} and in vivo fluorescence imaging \cite{Rao07}. RROC extends this approach to cold and ultracold molecules for robust high-fidelity readout in any setting for the advancement of molecular quantum science.

\begin{acknowledgments}
We thank P. L. Gould and the late E. E. Eyler for helpful discussions that led to the idea of resonance Raman optical cycling and thank A. E. Barentine for his comments on the manuscript. We gratefully acknowledge support from the NSF (CAREER Award No. 1848435) and the University of Connecticut, including a Research Excellence Award from the Office of the Vice President for Research.
\end{acknowledgments}

\vspace{0.2cm}
\noindent$^{\dagger}$ jamie.shaw@uconn.edu\\

\noindent$^{*}$ daniel.mccarron@uconn.edu

\bibliography{thebib}

\end{document}